\documentclass[12pt]{article}
\usepackage{amsfonts}
\usepackage{amssymb}

\def\hybrid{\topmargin -20pt    \oddsidemargin 0pt
        \headheight 0pt \headsep 0pt
        \textwidth 6.35in       
        \textheight 9.25in       
        \marginparwidth .875in
        \parskip 5pt plus 1pt   \jot = 1.5ex}

\hybrid

\def\baselinestretch{1.2}

\catcode`\@=11

\def\marginnote#1{}
\newcount\hour
\newcount\minute
\newtoks\amorpm
\hour=\time\divide\hour by60
\minute=\time{\multiply\hour by60 \global\advance\minute by-\hour}
\edef\standardtime{{\ifnum\hour<12 \global\amorpm={am}%
        \else\global\amorpm={pm}\advance\hour by-12 \fi
        \ifnum\hour=0 \hour=12 \fi
        \number\hour:\ifnum\minute<10 0\fi\number\minute\the\amorpm}}
\edef\militarytime{\number\hour:\ifnum\minute<10 0\fi\number\minute}

\def\draftlabel#1{{\@bsphack\if@filesw {\let\thepage\relax
   \xdef\@gtempa{\write\@auxout{\string
      \newlabel{#1}{{\@currentlabel}{\thepage}}}}}\@gtempa
   \if@nobreak \ifvmode\nobreak\fi\fi\fi\@esphack}
        \gdef\@eqnlabel{#1}}
\def\@eqnlabel{}
\def\@vacuum{}
\def\draftmarginnote#1{\marginpar{\raggedright\scriptsize\tt#1}}

\def\draft{\oddsidemargin -.5truein
        \def\@oddfoot{\sl preliminary draft \hfil
        \rm\thepage\hfil\sl\today\quad\militarytime}
        \let\@evenfoot\@oddfoot \overfullrule 3pt
        \let\label=\draftlabel
        \let\marginnote=\draftmarginnote
   \def\@eqnnum{(\theequation)\rlap{\kern\marginparsep\tt\@eqnlabel}%
\global\let\@eqnlabel\@vacuum}  }

\def\preprint{\twocolumn\sloppy\flushbottom\parindent 2em
        \leftmargini 2em\leftmarginv .5em\leftmarginvi .5em
        \oddsidemargin -.5in    \evensidemargin -.5in
        \columnsep .4in \footheight 0pt
        \textwidth 10.in        \topmargin  -.4in
        \headheight 12pt \topskip .4in
        \textheight 6.9in \footskip 0pt
        \def\@oddhead{\thepage\hfil\addtocounter{page}{1}\thepage}
        \let\@evenhead\@oddhead \def\@oddfoot{} \def\@evenfoot{} }

\def\numberbysection{\@addtoreset{equation}{section}
        \def\theequation{\thesection.\arabic{equation}}}

\def\underline#1{\relax\ifmmode\@@underline#1\else
        $\@@underline{\hbox{#1}}$\relax\fi}

\def\titlepage{\@restonecolfalse\if@twocolumn\@restonecoltrue\onecolumn
     \else \newpage \fi \thispagestyle{empty}\c@page\z@
        \def\thefootnote{\fnsymbol{footnote}} }

\def\endtitlepage{\if@restonecol\twocolumn \else \newpage \fi
        \def\thefootnote{\arabic{footnote}}
        \setcounter{footnote}{0}}  

\catcode`@=12
\relax

\def\figcap{\section*{Figure Captions\markboth
        {FIGURECAPTIONS}{FIGURECAPTIONS}}\list
        {Figure \arabic{enumi}:\hfill}{\settowidth\labelwidth{Figure
999:}
        \leftmargin\labelwidth
        \advance\leftmargin\labelsep\usecounter{enumi}}}
 \relax
\def\tablecap{\section*{Table Captions\markboth
        {TABLECAPTIONS}{TABLECAPTIONS}}\list
        {Table \arabic{enumi}:\hfill}{\settowidth\labelwidth{Table
999:}
        \leftmargin\labelwidth
        \advance\leftmargin\labelsep\usecounter{enumi}}}
 \relax
\def\reflist{\section*{References\markboth
        {REFLIST}{REFLIST}}\list
        {[\arabic{enumi}]\hfill}{\settowidth\labelwidth{[999]}
        \leftmargin\labelwidth
        \advance\leftmargin\labelsep\usecounter{enumi}}}
 \relax
\makeatletter
\newcounter{pubctr}
\def\publist{\@ifnextchar[{\@publist}{\@@publist}}
\def\@publist[#1]{\list
        {[\arabic{pubctr}]\hfill}{\settowidth\labelwidth{[999]}
        \leftmargin\labelwidth
        \advance\leftmargin\labelsep
        \@nmbrlisttrue\def\@listctr{pubctr}
        \setcounter{pubctr}{#1}\addtocounter{pubctr}{-1}}}
\def\@@publist{\list
        {[\arabic{pubctr}]\hfill}{\settowidth\labelwidth{[999]}
        \leftmargin\labelwidth
        \advance\leftmargin\labelsep
        \@nmbrlisttrue\def\@listctr{pubctr}}}
 \relax
\makeatother
\newskip\humongous \humongous=0pt plus 1000pt minus 1000pt

\newif\ifdtup

\relax

\def\be{\begin{equation}}
\def\ee{\end{equation}}
\def\ba{\begin{eqnarray}}
\def\ea{\end{eqnarray}}

\def\del{\partial}

\def\a{\alpha}

\def\d{\delta}

\def\e{\epsilon}

\def\th{\theta}

\def\m{\mu}

\def\s{\sigma}

\def\qq{\qquad}

\def\IR{\relax{\rm I\kern-.18em R}}

\def \ha {{1\over 2}}

\def \ov {\over}

\def\IR{\relax{\rm I\kern-.18em R}}
\def\inv{^{\raise.15ex\hbox{${\scriptscriptstyle -}$}\kern-.05em 1}}

\begin{document}

\renewcommand{\theequation}{\thesection.\arabic{equation}}

\newcommand{\beq}{\begin{equation}}
\newcommand{\eeq}[1]{\label{#1}\end{equation}}
\newcommand{\ber}{\begin{eqnarray}}
\newcommand{\eer}[1]{\label{#1}\end{eqnarray}}
\newcommand{\eqn}[1]{(\ref{#1})}
\begin{titlepage}
\begin{center}

\hfill hep--th/0205006\\
\hfill April 2002\\
\vskip .6in

{\Large \bf PP-waves and logarithmic conformal field theories}

\vskip 0.6in

{\bf Ioannis Bakas}$^1$\phantom{x} and\phantom{x}
 {\bf Konstadinos Sfetsos}$^2$
\vskip 0.1in

${}^1\!$
Department of Physics, University of Patras\\
26500 Patras, Greece\\
{\footnotesize{\tt bakas@ajax.physics.upatras.gr}}

\vskip .2in

${}^2\!$
Department of Engineering Sciences, University of Patras\\
26110 Patras, Greece\\
{\footnotesize{\tt sfetsos@mail.cern.ch, des.upatras.gr}}\\

\end{center}

\vskip .6in

\centerline{\bf Abstract}
We provide a world-sheet interpretation to the plane wave limit of a large
class of exact supergravity backgrounds in terms of
logarithmic conformal field theories. As an illustrative
example, we consider the two-dimensional conformal field theory
of the coset model $SU(2)_N/U(1)$ times a free time-like boson $U(1)_{-N}$,
which admits a space-time interpretation as a three-dimensional plane
wave solution by taking a correlated limit \`a la Penrose. We show that
upon a contraction of Saletan type, in which the parafermions
of the compact coset model are combined with the free time-like boson,
one obtains
a novel logarithmic conformal field theory with central charge $c=3$.
Our results are motivated at the classical level using
Poisson brackets of the fields,
but they are also explicitly demonstrated at the quantum level using
exact operator product expansions.
We perform several computations in this theory including the evaluation
of the four-point functions involving primary fields and their logarithmic
partners, which are identified. We also
employ the extended conformal symmetries of
the model to construct an infinite number of logarithmic operators.
This analysis can be easily generalized to other exact conformal field theory
backgrounds with a plane wave limit in the target space.

\noindent

\vskip .4in
\noindent

\end{titlepage}
\vfill
\eject

\def\baselinestretch{1.2}

\baselineskip 20pt

\section{Introduction}
\setcounter{equation}{0}

Searching for exact conformal field theories with interesting
space-time interpretation in string theory, as in black hole physics and cosmology,
has been a
very active area of research in recent years (see, for instance,
\cite{Witten}-\cite{Johnson}).
The advantages of having an exact description are rather obvious, since
one can in principle perform a quantum mechanical stringy investigation of
gravitational effects and deal with singularities and other important
issues that appear in classical gravitational theories.
Nevertheless, the exact description of a gravitational background in string
theory does not necessarily guarantee that the corresponding
conformal field theory will be easy
to solve.
The reason is that the resulting conformal field theories can be quite complicated,
as they are
based on non-compact groups that are difficult to analyze even at the Lie
algebra level \cite{nonco}-\cite{Gawedzki}.
Moreover, there are many other backgrounds of great physical interest that
are still awaiting an exact conformal field theory description. While many
technical questions concerning the applications of string theory to gravitational
problems are still remaining open to this day, any new progress is particularly
welcome.

Plane waves arise as classical solutions to theories of gravity
and share the attractive feature that they depart from
the trivial flat space solution in the most controllable possible manner.
This is essentially due to the existence of a covariantly constant
null Killing vector, which, as it turns out,
guarantees that curvature effects are kept to a minimum.
Plane waves, being relatively simple solutions, are particularly easy to
treat in various computations, which are usually performed
on gravitational backgrounds at the semi-classical level.
Therefore, the plane wave solutions,
although might not be as interesting as, for instance,
black hole and other solutions, they have the
advantage of being simple enough to
analyze in depth, yet without being completely trivial
as flat space-time.
Combining with the previous remarks, it is then natural to expect that
the conformal field theories (CFT)
corresponding to plane wave backgrounds will be relatively easy to study
and in fact prove that they can be completely solvable.

Two distinct categories of exact CFT with
the space-time interpretation of plane waves were constructed in
recent years.
The first category comprises of current algebra theories corresponding
to WZW models based on non-semi-simple groups. The
prototype example is the four-dimensional plane wave solution of \cite{NaWi},
which was further analyzed using CFT techniques in \cite{KiKouLu}.
This model can also be obtained using a limiting procedure
starting from the action
for the WZW model for $SU(2)_N\times U(1)_{-N}$, where the last factor
represents a time-like free boson playing
the r\^ole of time \cite{sfe1}. This procedure is actually a limit of
Penrose type that was first introduced in general relativity
\cite{Pen}, but here it also involves a NS--NS two-form.
The limit of WZW models involves taking the level $N$ of the current
algebra very large, but in such a correlated way with a rescaling
of the variables, that the resulting geometry
is a plane wave, instead of flat space. This particular limiting
procedure works out as a contraction of Saletan type for the underlying
current algebra.
Using these methods, a large
class of models have been constructed in \cite{Olive,sfe2} and
various aspects of current algebras based on non-semisimple groups
were subsequently investigated \cite{Mohammedi}-\cite{Forgacs}.
Note for completeness that the Lie algebra analog of this procedure
is fairly known
in the mathematical literature
(see, for instance, \cite{Gilmore}). We only emphasize here that the Saletan
contraction is different from
the Inonu--Wigner contraction, which leads to flat space-time,
instead of a plane wave solution, and hence to a
trivial CFT.

The second category of plane wave backgrounds with an exact
CFT interpretation in string theory, comprises of
coset theories corresponding to
gauged WZW models based on non-semisimple
groups. These can also be obtained, as it turns out in all known cases,
by performing the same limiting procedure of Penrose type;
illustrative
examples of such models can be found in \cite{sfe1,Antoniadis,sfetse}.
However, a detailed analysis of such
theories is still missing at the CFT level and it is
the purpose of this paper to initiate such investigation.
We will find that the resulting theories belong to the class of
so-called Logarithmic Conformal Field Theories
(LCFT) \cite{saleur,Gur} which have been the subject of
intense research over the past few years for various diverse reasons
\cite{BK}-\cite{persis}, starting with condensed matter
physics \cite{gardy,saleur2};
for a recent review
of the subject see \cite{Flrev} and references therein.
Curiously, our present work is the first
in the literature to provide an example of a LCFT
with clear space-time interpretation.

We will illustrate the main idea of our construction
by focusing on a three-dimensional plane wave background,
which is constructed from the metric and dilaton fields corresponding
to the direct product theory $(SU(2)_{N}/U(1)) \times U(1)_{-N}$.
We have, in particular,
\ba
&&{1\ov N} ds^2 =-  dt^2 +   d\th^2 + \cot^2\th d\phi^2  \ ,
\nonumber\\
&& e^{-2\Phi}= {e^{-2 \Phi_0}} \sin^2\th\ .
\label{metdil}
\ea
Consider next the change of variables from $(t,\th,\phi)\to (u,v,x)$,
given by
\be
\th=\e v + u \ ,\qq t=u\ ,\qq \phi=\sqrt{\e}\ x\ ,\qq N={\a\ov \e}\ ,
\label{liimi}
\ee
where $\a,\e$ are two parameters, followed by the limit $\e\to 0$;
$\alpha$ remains finite and fixed through the limiting process.
Then, the resulting gravitational background is given by
\ba
&& {1\ov \a} ds^2  = 2 dudv + \cot^2u dx^2  \ ,
\nonumber\\
&& e^{-2\Phi}= {e^{-2 \Phi_0}} \sin^2 u\ .
\label{metdli}
\ea
It represents the three-dimensional plane wave solution of
interest, which has also been constructed in \cite{sfetse} in the same
way.

The problem we would like to address in this paper
is to find the nature of the exact
CFT corresponding to the gravitational background
\eqn{metdli}. Along the way, we will also be interested in what happens
to the symmetries
of the original model \eqn{metdil} under the plane wave limit \eqn{liimi}.
The relevant discussion will naturally involve the time-like boson
corresponding to the factor $U(1)_{-N}$, but also the compact
parafermions of the coset model $SU(2)_N/U(1)$ \cite{paraf}, which
are the natural chirally conserved objects for this exact CFT.
In section 2, for the benefit of many readers, we will therefore
review the basic elements of the parafermionic operator algebra,
following \cite{paraf}, which
can nevertheless be skipped by the experts. In section 3, we consider
the classical counterpart of the parafermion currents that were introduced
systematically in \cite{claspara}. As we will see later,
they admit a representation
in terms of the metric variables
appearing in \eqn{metdil} and they obey
Poisson bracket relations as classical counterpart of the parafermionic
operator algebra. The limiting procedure \eqn{liimi}
will be naturally manifested at the Poisson algebra level and it will
provide a guide on how to perform a similar limit quantum mechanically.
Hence, the emergence of an exact LCFT will also be evident already at
the classical level.
In section 4, we perform the limiting procedure at the quantum level, using
operator product expansions among the simplest operators of the model,
and we explicitly show the emergence of a LCFT,
as has been advertized. We also compute 2- and 4-point functions of operators
that are logarithmic partners of each other and include a brief
review of the relevant aspects of logarithmic conformal field theories.
In section 5, we obtain as byproduct a free field realization
of the elementary logarithmic partner fields. We also
employ the extended conformal symmetries of the model
to construct an infinite number of logarithmic operators. However, our
analysis is not complete in the general case as we have not yet studied the
representation theory of the limiting model in great detail. More results
in this direction will be reported elsewhere.
Finally, in section 6, we present the conclusions and briefly discuss
some future directions.

It is also worth mentioning that although our present results are limited
to a very simple background with an exact conformal field theory description,
more general models can be tackled in a similar fashion. The two most important
aspects of our work are first the world-sheet interpretation of the plane
wave limit in supergravity theory and second the construction of a novel
new class of logarithmic conformal field theories that comes as a bonus in
this interpretation.

\section{Parafermion operator algebra}
\setcounter{equation}{0}
In the following, we turn to the
parafermion algebra that arises in the quantum
mechanical treatment of the conformal field theory $SU(2)_N/U(1)$ for any integer
value of the level $N$ and prepare to discuss carefully its large $N$ limits that are
appropriate for the extended coset model $(SU(2)_{N}/U(1)) \times U(1)_{-N}$.
We review the basic elements of the parafermionic operator algebra,
following \cite{paraf}, and provide a brief summary of their correlation
functions that will be used in the sequel. In the next section
we will also
consider the classical analogue of the parafermion currents, together
with their
commutation relations with respect to Poisson brackets,
in order to motivate (among
other things) appropriate field redefinitions and the new
algebraic structures that
emerge by taking $N$ to infinity. This will make it possible to extract the
basic information on the world-sheet in terms of the target space coordinates
and examine the different contractions of the resulting
gravitational background,
like its plane wave (Penrose) limit.

Recall that in ordinary conformal field theories with $Z_N$ symmetry, as in the coset
model $SU(2)_N/U(1)$, there exists a set of $2N-1$ parafermion fields $\psi_l(z)$ and
$\bar{\psi}(\bar{z})$ with $l = 0, 1, \cdots , N-1$ and $\psi_0 = 1 = \bar{\psi}_0$
corresponding to the identity operator, which are chirally conserved, i.e.,
$\bar{\partial} \psi_l = 0$ and $\partial \bar{\psi}_l = 0$.
The parafermion fields are semi-local generalizing the usual
fermion fields of the simplest $Z_2$ model to any other integer value $N$ and have
fractional conformal dimension $d_l$ depending on $N$. First note that the fields
$\psi_l(z)$ and $\bar{\psi}_l(\bar{z})$ have $Z_N \times \tilde{Z}_N$ charges
$(l, l)$ and $(l, -l)$ respectively and as a result one picks up a phase
by encircling one parafermion field around another inside a correlation function.
Since any two fields with charges $(p, q)$ and $(p^{\prime}, q^{\prime})$
yield a phase factor $\Omega = {\rm exp}(2 \pi i \theta)$, where
$\theta = -(pq^{\prime} + p^{\prime}q)/N$, the chiral fields $\psi_l$ and
$\psi_{l^{\prime}}$ have a mutual locality exponent $\theta = -2ll^{\prime}/N$.
For $N=2$, we have the usual fermions in two dimensions with $l=l^{\prime} = 1$
and $\Omega =1$, whereas $\theta$ is non-integer for $N >2$; local fields with
$\Omega = 1$ are recovered again in the large $N$ limit, as the parafermions turn into
bosons. Moreover, the operator product expansion of two parafermions assumes the
general form
\be
\psi_l (z) \psi_{l^{\prime}} (w) = {C_{l, l^{\prime}} \over (z-w)^{d_l + d_{l^{\prime}}
- d_{l + l^{\prime}}}} \sum_{n=0}^{\infty} (z-w)^n \Psi_{l+l^{\prime}}^{(n)} (w) ~,
\label{gope}
\ee
where $\Psi_{l + l^{\prime}}^{(n)}$ are appropriate fields with charges
$(l+l^{\prime}, l+l^{\prime})$ and $C_{l, l^{\prime}}$ are suitable structure
constants. The leading term in the power series expansion, $\Psi_{l+l^{\prime}}^{(0)}$,
coincides with the parafermion field $\psi_{l+l^{\prime}}$, whereas more complicated
terms follow to higher orders in the operator product expansion. The anti-holomorphic
sector of the theory that describes the operator product expansion of the
remaining parafermion fields $\bar{\psi}_l(\bar{z})$ among themselves is identical
to \eqn{gope}
using the anti-holomorphic coordinates $\bar{z}$ and $\bar{w}$ instead of $z$ and
$w$, and it will be omitted from now on.

Taking into account the mutual locality exponent of the fields
$\psi_l$ and $\psi_{l^{\prime}}$, one easily obtains the following relation for
the conformal dimension of the parafermion fields:
\be
d_l + d_{l^{\prime}} - d_{l+l^{\prime}} = 2ll^{\prime}/N ~~~~ {\rm mod} Z ~.
\ee
The conformal dimension of the $N$ chiral fields $\psi_l(z)$ that solves this equation
and defines the parafermion operator algebra for $Z_N$ symmetric theories is
\be
d_l = d_{N-l} = {l(N-l) \over N} ~; ~~~~ l = 0, 1, 2, \cdots , N-1 ~,
\ee
where we also impose the hermiticity relation $\psi_l^{\dagger}(z) = \psi_{N-l}(z)$
among its generators. Then, the operator product expansion of the parafermion fields
assumes the following complete form:
\ba
\psi_l(z) \psi_{l^{\prime}}(w) & = & {C_{l, l^{\prime}} \over (z-w)^{2ll^{\prime}/N}}
\left(\psi_{l+l^{\prime}}(w) + {\cal O}(z-w) \right) ~; ~~~~ {\rm for} ~~~
l+l^{\prime} <N ~, \nonumber\\
\psi_l(z) \psi_{l^{\prime}}^{\dagger}(w) & = & {C_{l, N-l^{\prime}} \over
(z-w)^{2l(N-l^{\prime})/N}}
\left(\psi_{l-l^{\prime}}(w) + {\cal O}(z-w) \right) ~; ~~~~ {\rm for} ~~~
l^{\prime}<l ~, \label{paat}\\
\psi_l(z) \psi_l^{\dagger}(w) & = & {1 \over
(z-w)^{2l(N-l)/N}}
\left(1 + {2d_l \over c} (z-w)^2 T(w) + {\cal O}(z-w)^3 \right). \nonumber
\ea
The parameter $c$ is the central charge
of the Virasoro algebra generated by the
stress-energy tensor of the model, which is given by
\be
c = 2{N-1 \over N+2}\ ,
\ee
for all integer $N$. Moreover, the structure constants $C_{l, l^{\prime}}$ are determined
by the associativity property of the operator product expansion and they turn out to be
\be
C_{l, l^{\prime}}^2 = {\Gamma(N-l+1) \Gamma(N-l^{\prime} +1) \Gamma(l+l^{\prime}+1)
\over \Gamma(l+1) \Gamma(l^{\prime} +1) \Gamma(N+1) \Gamma(N-l-l^{\prime} +1)}
\ .
\ee
One clearly has $C_{l, N-l} =1$.

We add for completeness the operator product expansion of the
stress-energy tensor $T(z)$
of the model with the parafermion currents,
\be
T(z) \psi_l(w) = {d_l \over (z-w)^2} \psi_l(w) + {1 \over z-w} \partial \psi_l (w)
+ {\cal O}(1) ~,
\ee
which state that they are primary fields with conformal dimension $d_l$, and the operator
product expansion with the stress-energy tensor itself,
\be
T(z) T(w) = {c \over 2(z-w)^4} + {2 \over (z-w)^2} T(w) + {1 \over z-w} \partial T(w)
+ {\cal O}(1) ~,
\ee
which yields the Virasoro algebra with central charge $c$, as it is required.

The correlation functions among parafermion fields can be computed using the operator
product expansions above. For instance, the 2-point function
\be
\langle \psi_l(z) \psi_{l^{\prime}}^{\dagger}(w)\rangle
= {\delta_{l, l^{\prime}} \over (z-w)^{2d_l}} \ ,
\ee
follows immediately and provides the standard
normalization of the parafermions. Higher
order correlation functions
also exhibit singularities with power law behaviour, which
can be determined recursively using the
structure of the operator algebra. Thus, a
$2n$-point parafermion correlation function is
related to the $(2n-2)$-point function as in
\ba
& & \langle\psi_1(z_1) \cdots \psi_1(z_n)
\psi_1^{\dagger}(w_1) \cdots \psi_1^{\dagger}(w_n)\rangle =
\prod_{i=2}^n {1 \over (z_1 - z_i)^{2/N}}
\prod_{j=1}^n (z_1 - w_j)^{2/N} \times \nonumber\\
& & \sum_{k=1}^n \left( {1 \over (z_1 - w_k)^2} + {2 \over N(z_1 - w_k)}
\left(\sum_{l=2}^n {1 \over w_k -z_l} - \sum_{m \neq k}^n {1 \over w_k -w_m}
\right) \right)
\prod_{q=2}^n (z_q - w_k)^{2/N} \times \nonumber\\
& & \prod_{p=1}^{k-1} {1 \over (w_p - w_k)^{2/N}}
\prod_{r=k+1}^n {1 \over (w_k - w_r)^{2/N}}
\langle \psi_1(z_2) \cdots \psi_1(z_n)
\psi_1^{\dagger}(w_1) \cdots \hat{\psi}_1^{\dagger}(w_k)
\cdots \psi_1^{\dagger}(w_n)\rangle \ ,
\nonumber
\\ &&
\ea
where $\hat{\psi}_1^{\dagger}(w_k)$ means that this factor has been removed by taking
the operator product expansion with $\psi_1(z_1)$.
Although general correlation functions can be obtained in closed form by straightforward
but cumbersome iteration of the formula, it will be
sufficient for our purposes to consider only the exact form of the 4-point parafermion
function, which turns out to be
\be
\langle \psi_1(z_1) \psi_1^{\dagger}(z_2) \psi_1(z_3)
\psi_1^{\dagger}(z_4)\rangle  =
\left({z_{12} z_{14} z_{34} z_{23} \over z_{13} z_{24}}\right)^{2/N}
\left({1 \over z_{12}^2 z_{34}^2} \left(1 + {2 \over N} {z_{12} z_{34} \over
z_{23}z_{24}} \right) + (z_2 \leftrightarrow z_4) \right)
\label{fptf}
\ee
where $z_{ij} = z_i - z_j$ stands for simplicity.

We conclude this section with a result that has been known for some time
\cite{baki,prs}. Recall
that the parafermion currents become bosonic in the limit
$N \rightarrow \infty$, and as a result the
various operator product expansions of
the parafermion fields as well as their correlation functions become identical to
those of two real free bosons (or equivalently a complex free boson) in two dimensions.
In particular, one obtains an infinite selection of fields $\psi_l(z)$ and
$\psi_l^{\dagger}(z)$ with $l = 0, 1, 2, \cdots,$ having integer conformal dimension
$d_l = l$, which can be represented in terms of a complex scalar field $\phi(z)$ as
\be
\psi_l(z) = :\left(i\partial \phi(z)\right)^l: ~, ~~~~~
\psi_l^{\dagger}(z) = :\left(-i\partial \bar{\phi}(z)\right)^l:
\ee
and can be used to obtain a bosonic field realization of all subleading terms in
the operator product expansion of the $Z_{\infty}$ parafermion currents. In other words,
$\psi_1(z)$ and $\psi_1^{\dagger}(z)$ are the two basic bosonic currents, whereas the
remaining higher $Z_{\infty}$ parafermions belong in their enveloping algebra.
Then, the central charge of the Virasoro algebra of the $SU(2)_N/U(1)$ coset
model, which is given by $c = 2(N-1)/(N+2)$,
assumes its classical value $c=2$ as $N \rightarrow \infty$.
On the other hand, the
free boson model $U(1)_{-N}$ has a current algebra generated by the chiral fields
$J_0(z)$ (and $\bar{J}_0(\bar{z})$ for the anti-holomorphic sector)
with conformal dimensions $(1, 0)$ and $(0, 1)$
respectively. The operator product expansion of the holomorphic currents is
\be
J_0(z) J_0(w) = -{N \over 2(z-w)^2} + {\cal O}(1)
\ee
and also has a well defined large $N$ limit obtained by rescaling the
components of the $U(1)$ current by $\sqrt{N}$. The central charge of the Virasoro
algebra is $c=1$ in this case, and therefore the total central charge of the product
model $(SU(2)_{N}/U(1)) \times U(1)_{-N}$
becomes 3 in the large $N$ limit, thus describing
three free bosons in total. In accordance with this limit the corresponding
background metric in \eqn{metdil} becomes flat and the dilaton constant.

It is important to realize that there is an alternative
large $N$ limit of the same product model that will be
taken in a correlated way between the two factors, which
yields an inequivalent theory having logarithmic structure in
its correlation functions. This logarithmic CFT will be
discussed later in all detail, but the central charge will
remain $c=3$ irrespective of the limiting procedure.


\section{Classical parafermion currents}
\setcounter{equation}{0}

In this section we consider the classical analogue of the parafermion
currents, together with their
commutation relations with respect to Poisson brackets,
in order to motivate the appropriate field redefinitions
and the new algebraic structures that
emerge by taking $N\to \infty$. In terms of the target space variables
they are represented as \cite{claspara}
\ba
&&
\psi_1= \sqrt{N} (\del_+\th
 + i \cot\th \del_+ \phi) e^{-i(\phi-\int \cot^2\th \del_+\phi)}\ ,
\nonumber\\
&& \psi_1^\dagger= \sqrt{N}
(\del_+\th - i \cot\th \del_+ \phi) e^{+i(\phi-\int \cot^2\th \del_+\phi)}
\label{parra}\ .
\ea
We chose appropriate overall normalization factors so that they are the same
for both classical and quantum parafermion currents.

It can be easily shown, using the equations of motion that follow from the
two-dimensional $\s$-model action corresponding to the metric \eqn{metdil},
that they are chirally conserved, i.e., $\del_- \psi_1=\del_- \psi_1^\dagger=0$.
It is also well known that they obey the equal ``time''
(with respect to $\s_-$,
say) Poisson bracket algebra \cite{claspara}
\ba
&&
\{\psi_1,\psi_1\}={\pi \ov 2N} \e(\s_+-\s'_+)
\psi_1(\s_+) \psi_1(\s'_+) \ ,
\nonumber\\
&& \{\psi_1^\dagger,\psi_1^\dagger\}={\pi \ov 2N} \e(\s_+-\s'_+)
\psi_1^\dagger(\s_+) \psi_1^\dagger(\s'_+) \ ,
\label{algpaa}\\
&& \{\psi_1,\psi_1^\dagger\}=-{2\ov \pi} \d'(\s_+-\s'_+) -
{\pi \ov 2N} \e(\s_+-\s'_+)
\psi_1^\dagger(\s_+) \psi_1(\s'_+) \ ,
\nonumber
\ea
where one assumes in writing Poisson brackets that the first parafermion
current is evaluated at $\s_+$ and
the second one at $\s'_+$, whereas $\s_-$ is common to both of them. Also
we have used the antisymmetric step function defined as
$\e(\s_+-\s'_+)=1 (-1)$ if $\s_+>\s'_+$ ($\s_+<\s'_+$) and
$\d'$ is the derivative of the usual $\d$-function.
In addition to \eqn{algpaa}, we have the Poisson bracket
corresponding to the time-like boson,
\be
\{J_0,J_0\}={N\ov \pi} \d'(\s_+-\s'_+) \ ,
\label{allbo}
\ee
where $J_0=-N \del_+ t$.

Taking the limit \eqn{liimi}, we find the following
interesting behaviour for the parafermion current:
\be
\sqrt{N} \psi_1
= \ha \Psi +{1\ov 2\e} \Phi + {i\ov \sqrt{\e}} P + {\cal O}(\sqrt{\e}) \,
\label{ppio}
\ee
whereas for $\psi_1^\dagger$ we get its complex conjugate. Then, the
real fields $\Psi$, $\Phi$ and $P$ can be represented in terms of
the variables of the background \eqn{metdli} as
\ba
&& \Psi=\a ( 2 \del_+v + 2 \cot u \del_+x - A^2 \del_+u)\ ,
\nonumber\\
&& \Phi = 2 \a \del_+ u  \ ,
\label{pspha}\\
&& P = \a(\cot u \del_+ x - A \del_+u )\ ,
\nonumber
\ea
where $A=x-\int \cot^2 u \del_+x $.
These fields are chirally conserved on-shell by construction,
but one may nevertheless check this explicitly.
We note for consistency that the same limiting procedure of the extra $U(1)$
generator $J_0$ yields an expression for the field $\Phi$ which is
identical to the one appearing in \eqn{pspha}; thus, our limiting procedure
is sensible and we may proceed further with the calculations.
The Poisson brackets for $\Phi,\Psi$ and $P$ are easily
found using \eqn{algpaa}, \eqn{allbo} and \eqn{ppio}.
The result reads
\ba
&&\{\Phi,\Phi\}=0\ ,
\nonumber\\
&&\{\Phi,\Psi\}= -{2\a\ov \pi} \d'(\s_+-\s'_+)\ ,
\nonumber\\
&&\{\Psi,\Psi\}= -{\pi\ov 2 \a} \e(\s_+-\s'_+)
P(\s_+)P(\s'_+)\ ,
\nonumber\\
&&\{P,P\}= -{\a\ov \pi}\d'(\s_+-\s'_+) -{\pi\ov 8\a} \e(\s_+-\s'_+)
\Phi(\s_+)\Phi(\s'_+)\ ,
\label{ppff}\\
&&\{\Phi,P\}= 0\ ,
\nonumber\\
&&\{\Psi,P\} = {\pi\ov 4 \a} \e(\s_+-\s'_+) P(\s_+)\Phi(\s_+')\ .
\nonumber
\ea

Note that unlike the
Poisson bracket algebra of the classical parafermions \eqn{algpaa},
which is extended non-trivial at the quantum level in \eqn{paat} by receiving
$1/N$-corrections to all orders, the algebra \eqn{ppff}
is actually exact to all orders in $\a$ in the
world-sheet perturbation theory.
There are several ways to see this as, for instance,
the solution \eqn{metdli} does not receive any
quantum corrections in string perturbation theory on the world-sheet
due to its plane wave nature \cite{Amati,Horowitz}.
This is easily argued using the fact that the constant $\a$ appearing in
\eqn{metdli}, which sets the scale in perturbation theory,
can be set equal to any value by an appropriate rescaling
of the variables $v$ and $x$. Hence, it is natural to
expect that the classical symmetries
of this model will also be exact symmetries at
the quantum level.  This observation is also in agreement
with the fact that the constant $\a$ appearing in the algebra
\eqn{ppff} can be absorbed by obvious simultaneous rescaling of
the currents $\Phi$ and
$P$ in an obvious way. Keeping these remarks in mind, we are
setting $\a=1$ from now in
the rest of the paper.

\section{Correlated large $N$ limit and LCFT}
\setcounter{equation}{0}

We return now to the operator product expansions of the
quantum operators of the
model $(SU(2)_{N}/U(1)) \times U(1)_{-N}$ and consider the large $N$
limit that is
obtained by forming linear combinations of the
parafermion currents of the $SU(2)_N/U(1)$ factor
with the $U(1)_{-N}$ currents.
We will find this appropriate for taking the plane wave limit
of the extended coset model as it has already been indicated by our
previous classical
analysis in terms of Poisson brackets.
In this context, we will arrive at a rather
surprising result stating that the correlated
large $N$ limit defines a logarithmic conformal
field theory for the model $(SU(2)_\infty/U(1))\times U(1)_{-\infty}$
with central charge $c=3$,
as for the ordinary large $N$ limit. The new feature here
is the appearance of logarithmic divergences in certain
correlation functions that are defined in this limit, and hence one can interpret the
plane wave limit of ordinary conformal field theories as
providing new examples of logarithmic
conformal field theories (LCFT), which have not been realized so far to the best of
our knowledge. Although our analysis is presently confined to the
coset model $(SU(2)_{N}/U(1)) \times U(1)_{-N}$,
generalizations can also be considered
to other conformal field theories as $N \rightarrow \infty$.

Following the lines of the classical analysis presented earlier, it is natural to
define the three Hermitian chiral operators
\ba
\Phi(z) & = & \epsilon \left({\sqrt{N} \over 2} (\psi_1 + \psi_1^{\dagger}) - J_0\right) ,
~~~~ \Psi(z) = {\sqrt{N} \over 2} (\psi_1 + \psi_1^{\dagger}) + J_0
~,\nonumber\\
P(z) & = & \sqrt{\epsilon} {\sqrt{N} \over 2i} (\psi_1 - \psi_1^{\dagger} )
\ea
instead of $\psi_1(z)$, $\psi_1^{\dagger}(z)$ and $J_0(z)$ whose $U(1)$ central charge
is $-N$; the non-vanishing 2-point functions
$\langle \psi_1(z) \psi_1^{\dagger}(w)\rangle $ and
$\langle J_0(z) J_0(w) \rangle $ are then normalized accordingly to 1 and
$-N/2$ respectively.
The parameter $\epsilon$ is defined as before,
i.e., $\epsilon = \alpha/N=1/N$. 
Note that the first two linear combinations do not have well-defined conformal
dimension for finite values of $N$, as
the weight is $1- 1/N$ for each one of the parafermion
currents and $1$ for the $U(1)$ current $J_0(z)$; however,
the dimensions match in the
large $N$ limit that will be taken shortly.
Also, in the limit $N \rightarrow \infty$,
the operator $\Phi(z)$ appears to scale to zero as $1/\sqrt{N}$,
the operator $\Psi(z)$
scales to infinity as $\sqrt{N}$, whereas $P(z)$ remains finite.
Finally, the stress-energy tensor
of the extended model is defined as usual,
\be
T(z) = T_{SU(2)/U(1)} + T_{U(1)}
\ee
and will be used to compute the operator product expansion
with each one of the
fields $\Phi(z)$, $\Psi(z)$ and $P(z)$. It is here that we
will first encounter
the characteristic behaviour of a LCFT upon taking the large $N$ limit.

We first compute the operator product expansion
\ba
&& T(z) \Psi(w) = {1 \over 2(z-w)^2}
\left(\left(1 - {1 \over N}\right) \left(\Psi(w)
+ {1 \over \epsilon} \Phi(w)\right) + \Psi(w)
- {1 \over \epsilon} \Phi(w)\right)
\nonumber\\
&&\phantom{xxxxxxxxxxx} + {\partial \Psi(w) \over z-w} + {\cal O}(1) \ ,
\ea
using the definition of the field $\Psi$ and rewriting the
right-hand side in terms
of $\Phi$ and $\Psi$. Although there is nothing
particular happening for finite values
of $N$, we observe that in the large $N$ limit ($\epsilon \rightarrow 0$)
the term which appears to order
$1/(z-w)^2$ becomes
\be
{1 \over 2}\lim_{\epsilon \rightarrow 0}
\left(\left(1 - {1 \over N}\right) \left(\Psi +
{1 \over \epsilon} \Phi\right) + \Psi
 - {1 \over \epsilon} \Phi \right) =
\Psi - {\Phi\ov 2}  ~.
\ee
Thus, we obtain in this limit the result
\be
T(z) \Psi(w) = {\Psi(w) - \Phi(w)/2 \over (z-w)^2} + {\partial \Psi(w)
\over z-w} + {\cal O}(1) ~,
\ee
which is a characteristic operator product expansion in
LCFT.
Before making this connection more precise, we also compute the operator
product expansion of the stress-energy
tensor with the remaining two fields under
study. We then obtain
\ba
&& T(z) \Phi(w) = {\epsilon \over 2(z-w)^2} \left(\left(1 - {1 \over N}\right)
\left(\Psi(w) +{1 \over \epsilon} \Phi(w) \right) - \left(\Psi(w) - {1 \over \epsilon}
\Phi(w) \right) \right)
\nonumber\\
&&\phantom{xxxxxxxxxxx}
+ {\partial F(w) \over z-w} + {\cal O}(1)\ ,
\ea
which yields in the large $N$ limit the result
\be
T(z) \Phi(w) = {\Phi(w) \over (z-w)^2} + {\partial \Phi(w) \over z-w} + {\cal O}(1) ~.
\ee
As for the field $P$, which has definite conformal dimension for all $N$, we obtain
in the large $N$ limit
\be
T(z) P(w) = {P(w) \over (z-w)^2} + {\partial P(w) \over z-w} + {\cal O}(1) ~,
\ee
stating that it is an ordinary primary field of weight 1.

Next, we compute the 2-point functions among the fields $\Psi$ and $\Phi$.
Using the 2-point functions of the parafermions and the $U(1)$ current $J_0$,
we first obtain
\be
\langle \Psi(z) \Psi(w)\rangle
= {N \over 2 (z-w)^2} \left((z-w)^{2/N} - 1 \right) ,
\ee
which is valid for all values of $N$. In taking the large $N$ limit,
we may expand
\be
(z-w)^{2/N} = 1 + {2 \over N} \ln(z-w) + {\cal O}\left({1 \over N^2}\right) ,
\ee
thus arriving at the result for the 2-point function
\be
\langle \Psi(z) \Psi(w)\rangle  = {\ln(z-w) \over (z-w)^2} ~,
\ee
which exhibits a logarithmic dependence on $z-w$ on top of the
usual power law behaviour of ordinary two-dimensional conformal field theories.
Proceeding along similar lines we compute the 2-point function between the
fields $\Phi$ and $\Psi$. In this case the logarithmic dependence cancels
as $N \rightarrow \infty$ and the final result in the large $N$ limit reads as
\be
\langle \Phi(z) \Psi(w) \rangle = {1 \over (z-w)^2} \ .
\ee
Finally, we obtain that
\be
\langle \Phi(z) \Phi(w)\rangle = 0 ~,
\ee
which is immediate and obvious in the large
$N$ limit as the field $\Phi$ scales like $1/\sqrt{N}$.
As for the 2-point function $\langle P(z)P(w)\rangle$,
the resulting expression is as
in ordinary conformal field theory for a primary field of weight 1, i.e.,
\be
\langle P(z)P(w)\rangle = {1 \over 2(z-w)^2} ~.
\ee
The remaining 2-point functions are zero.

We also write down for completeness
the operator product expansions of the various
fields obtained from the parafermion and the free boson operator algebras
after the limit is taken. We obtain, in particular,
\ba
&&\Psi(z)\Phi(w)= {1\ov (z-w)^2} + {\cal O}(1) \ ,
\nonumber\\
&&\Psi(z)\Psi(w)={\ln(z-w)\ov (z-w)^2} + 2 \ln(z-w) :P^2(w):
+\ha \ln^2(z-w) :\Phi^2(w): + {\cal O}(1)\ ,
\nonumber\\
&&\Psi(z) P(w)= - \ln(z-w) :(P\Phi)(w): + {\cal O}(1) \ ,
\label{oppe}\\
&&P(z)P(w)={1\ov 2 (z-w)^2} + \ha \ln(z-w) :\Phi^2(w): + {\cal O}(1) \ .
\nonumber
\ea

Summarizing the results we have obtained so far, it is rather amusing to note the
logarithmic dependence of the correlation functions that arises in this large $N$
limit, which signals that a LCFT is at work in
the plane wave (Penrose) limit of our exact conformal field theory background
$(SU(2)_{N}/U(1)) \times U(1)_{-N}$.
As such, it provides us with a world-sheet
interpretation of the plane wave limit in supergravity, which can also be generalized
to more complicated string backgrounds with an exact conformal field theory
description. This observation can also be used to provide us
with a new class of LCFT
beyond the class of examples that have been studied so
far. In this context, the field $\Phi$ is a primary field of weight 1, whereas the
field $\Psi$ constitutes its logarithmic partner; as for $P$, which is also a
primary field of weight 1, it has no logarithmic partner.

To be more precise, we
recall the bare facts of LCFT in two dimensions,
which involve a selection of primary fields $\{\Phi_h\}$ of conformal weight $h$
and their logarithmic partners $\{\Psi_h\}$. It is well known that in
a LCFT any two logarithmic partners transform
under a chiral conformal transformation
$z \rightarrow \tilde{z}(z)$ as doublet, which can be formally written as
\be
\left(\begin{array}{c}
\Phi(z) \\ \Psi(z)
\end{array} \right) = \left({\partial \tilde{z} \over \partial z}
\right)^{\left(\begin{array}{cc}
h & 0 \\ 1 & h
\end{array} \right)}
\left(\begin{array}{c}
\Phi(\tilde{z}) \\ \Psi(\tilde{z})
\end{array} \right)
\ee
and similarly for the anti-holomorphic sector.
The infinitesimal form of these transformations give rise to the particular
operator
product expansions of the stress-energy tensor with
the fields $\Phi$ and $\Psi$ that have been
encountered above, provided that $\Phi$ is rescaled by $-1/2$;
in our case, we also have $h=1$. Put it differently,
the states $|\Phi\rangle  = \Phi(z=0)| 0\rangle$ and $|\Psi\rangle
= \Psi(z=0)|0\rangle$, which
correspond to the two fields in question, satisfy the general relations
\ba
L_0 |\Phi\rangle  & = & h | \Phi\rangle \ ,
\nonumber\\
L_0 |\Psi \rangle  & = & h | \Psi\rangle  + | \Phi\rangle
\ea
and therefore the zero mode of the Virasoro algebra $L_0$ can no longer be diagonalized
in LCFT, as it is in ordinary conformal field theories.
Moreover, the 2-point function of any
two chiral primary fields vanishes,
\be
\langle \Phi_h(z) \Phi_{h^{\prime}}(w) \rangle = 0 ~,
\ee
while for correlators involving their logarithmic partners we have
the following general structure:
\ba
\langle \Phi_h(z) \Psi_{h^{\prime}}(w)\rangle & = &
\delta_{h,h^{\prime}} {A \over
(z-w)^{h + h^{\prime}}}\ ,
\nonumber\\
\langle \Psi_h(z) \Psi_{h^{\prime}}(w) \rangle & = & \delta_{h,h^{\prime}}
{B - 2A \ln(z-w) \over
(z-w)^{h + h^{\prime}}}
\ea
with $A$ and $B$ free constants; note that the constant $B$ can be
consistently set to zero by a field redefinition.
This is precisely the behaviour that we have also encountered
in the large $N$ limit of the 2-point functions of the fields
$\Psi$ and $\Phi$
with $h = h^{\prime} = 1$, provided that the free constants are set equal to
$A=-1/2$ and $B=0$, while $\Phi$ is also rescaled by $-1/2$.

Having established the logarithmic structure of the resulting large $N$ theory by
computing the relevant 2-point functions, we may now proceed to the
calculation of higher point correlation functions.\footnote{
We immediately conclude that all odd-point functions involving the
fields $\Phi,\Psi$ and $P$ of our model are zero.
This is due to the vanishing of correlators involving an
odd number of $\psi_1$ and $\psi_1^\dagger$ due to charge conservation and
similarly for correlators involving an odd number of $J_0$'s. Hence,
we will not mention the general form of the 3-point functions.
Also, all $2n$-point functions involving more that $n$ insertions of the
field $\Phi$ are zero in our case.
}
Recall that the form of the
4-point functions in LCFT can be easily obtained on general grounds;
we summarize below the general expressions
for the non-vanishing 4-point functions that involve at least
one primary field $\Phi$ and its logarithmic partner:
\ba
\langle \Phi(z_1) \Phi(z_2) \Phi(z_3) \Psi(z_4) \rangle & = &
\prod_{i<j} {z_{ij}}^{\mu_{ij}}
F^{(0)}(x), \nonumber\\
\langle \Phi(z_1) \Phi(z_2) \Psi(z_3) \Psi(z_4) \rangle & = &
\prod_{i<j} {z_{ij}}^{\mu_{ij}}
\left(F_{34}^{(1)}(x) - 2F^{(0)}(x) \ln(z_{34}) \right),  \nonumber\\
\langle \Phi(z_1) \Psi(z_2) \Psi(z_3) \Psi(z_4)\rangle & = &
\prod_{i<j} {z_{ij}}^{\mu_{ij}}
\Big[F_{234}^{(2)}(x) - \sum_{2 \leq i < j \leq 4}
\tilde{F}_{ij}^{(1)}(x) \ln(z_{ij})
\nonumber\\
& + & 2F^{(0)}(x) \left(\ln(z_{23}) \ln(z_{24}) + \ln(z_{23})
\ln(z_{34}) + \ln(z_{24}) \ln(z_{34}) \right)    \nonumber\\
& - & F^{(0)}(x) \left(\ln^2(z_{23}) + \ln^2(z_{24}) +
\ln^2(z_{34}) \right) \Big]
\label{geee}
\ea
and omit the general expression for the correlation function
$\langle\Psi(z_1) \Psi(z_2) \Psi(z_3) \Psi(z_4)\rangle$
that is rather lengthy. Note that
we used the abbreviations
\be
z_{ij} = z_i - z_j ~, \qq
\mu_{ij} = {1 \over 3}\sum_{k=1}^4 h_k - h_i - h_j   \ ,
\ee
in order to simplify the expressions,
where $h_i$ denotes the conformal dimension of the field inserted at any point $z_i$.
The coefficients $F^{(0)}(x)$, $F_{ij}^{(1)}(x)$ and $F_{ijk}^{(2)}(x)$ are functions
of the anharmonic ratio of the four points, $x= (z_{12} z_{34})/(z_{14} z_{32})$, as
it is required by conformal invariance, but whose explicit
form depend on the particular theory.
According to their definition, the functions
$F_{ij}^{(1)}(x)$ are all related to a
single function, say $F^{(1)}(x)$, by the transformation
\be
F_{34}^{(1)}(x) = F^{(1)}(x) ~, ~~~~ F_{23}^{(1)}(x) = F^{(1)} \left({1 \over 1-x}
\right) , ~~~~ F_{24}^{(1)}(x) = F^{(1)}(1-x) ~.
\ee
Finally, we used the abbreviation
\be
\tilde{F}_{ij}^{(1)}(x)
= F_{ik}^{(1)}(x) + F_{jk}^{(1)}(x) - F_{ij}^{(1)}(x)   \ ,
\label{abbr}
\ee
with $k$ corresponding to the index of the remaining third logarithmic
field in the correlation functions above.

We have calculated the corresponding
4-point functions of the operators $\Psi$ and $\Phi$ in the
$(SU(2)_{N}/U(1)) \times U(1)_{-N}$ model, using the 4-point function of
the parafermion operators $\langle\psi_1(z_1) \psi_1^{\dagger}(z_2)
\psi_1(z_3) \psi_1^{\dagger}(z_4)\rangle$ that has been given earlier.
Expanding the factor in \eqn{fptf},
\be
\left({z_{12} z_{14} z_{34} z_{23} \over z_{13} z_{24}}\right)^{2/N} =
1 + {2 \over N}\ln\left({z_{12} z_{14} z_{34} z_{23} \over z_{13} z_{24}}
\right) + {\cal O}\left({1 \over N^2}\right)
\ee
as a power series in $1/N$, we may subsequently take the large $N$ limit of the
correlation functions, which is appropriate for describing the plane wave limit
of the supergravity background corresponding to our extended coset model. As
for the 2-point functions, we also find here a remnant
logarithmic dependence of the
4-point functions as $N \rightarrow \infty$, which precisely fits the general
formulae of LCFT according to our
expectations. We spare the details of the
calculation and only give the result
\ba
&& \langle \Phi(z_1)\Phi(z_2)\Phi(z_3)\Psi(z_4)\rangle = 0 \ ,
\nonumber\\
&& \langle \Phi(z_1)\Phi(z_2)\Psi(z_3)\Psi(z_4)\rangle =
{1\ov z_{13}^2 z_{24}^2}+ {1\ov z_{14}^2 z_{23}^2}\ ,
\label{pppp}\\
&&  \langle \Phi(z_1)\Psi(z_2)\Psi(z_3)\Psi(z_4)\rangle =
{\ln z_{34}\ov z_{12}^2 z_{34}^2}
+ {\ln z_{24}\ov z_{13}^2 z_{24}^2}
+ {\ln z_{23}\ov z_{14}^2 z_{23}^2}
\ .
\nonumber
\ea

These correlators can be cast into the general form \eqn{geee}
using that $\m_{ij}=-2/3$, $\forall\ i,j=1,2,3,4$. We also have
for the three key functions that enter into the correlators
\ba
&& F^{(0)}=F^{(2)}_{234}=0\ ,
\nonumber\\
&& F^{(1)}= (1+x^2) \left(x\ov x-1\right)^{2/3}\ .
\ea
Of course, in order to match the overall normalization,
our field $\Phi$ has to be
rescaled by $-1/2$, as it was mentioned before.
From these expressions
we also compute the abbreviated functions in \eqn{abbr},
\ba
&& \tilde F^{(1)}_{34}= 2
\prod_{i<j=1}^4 z_{ij}^{2/3} {1\ov z_{12}^2 z_{34}^2}\ ,
\nonumber\\
&& \tilde F^{(1)}_{23}= 2
\prod_{i<j=1}^4 z_{ij}^{2/3} {1\ov z_{14}^2 z_{23}^2}\ ,
\\
&& \tilde F^{(1)}_{24}= 2
\prod_{i<j=1}^4 z_{ij}^{2/3} {1\ov z_{13}^2 z_{24}^2}\ .
\nonumber
\ea
Other correlation functions can also be computed in the large $N$ limit, if
it is necessary, using the same procedure. In particular, we can also
show that the 4-point function having only logarithmic fields assumes the form
given in the literature \cite{Flrev}.

\section{Free field representation and further results}
\setcounter{equation}{0}

We first consider the free field realization of the
parafermion operators, which follows from the standard free field representation
of the $SU(2)_N$ current algebra. Introducing two real free bosons,
\be
\langle \phi_i(z) \phi_j(w)\rangle = -\delta_{ij} \ln(z-w) ~;
~~~~ i, j = 1, 2~,
\ee
one may represent the elementary parafermion currents as
\ba
\psi_1 &=& {1 \over \sqrt{2}} \left( -\sqrt{1 + {2/N}}\ \partial \phi_1
+ i \partial \phi_2 \right) e^{+\sqrt{2/N}\ \phi_2}\ ,
\nonumber\\
\psi_1^\dagger &=& {1 \over \sqrt{2}} \left( \sqrt{1 + {2/N}}\ \partial \phi_1
+ i \partial \phi_2 \right) e^{-\sqrt{2/N}\ \phi_2}\ .
\ea
We will denote the time-like free boson for the $U(1)_{-N}$ factor
by $\phi_0$. It obeys
\be
\langle \phi_0(z) \phi_0(w)\rangle = \ln(z-w)\ .
\ee
The stress-energy tensor of the entire theory is
\be
T(z)= \ha (\del\phi_0)^2 -\ha (\del\phi_1)^2-\ha (\del\phi_2)^2
+{i\ov \sqrt{2(N+2)}} \del^2 \phi_1 \ .
\ee

Let us now consider the scalar field redefinition
\be
\phi_+ = \sqrt{1 \ov 2N} (\phi_0+\phi_1)\ ,\qq
\phi_- = \sqrt{N\ov 2} (\phi_0-\phi_1)\ ,\qq \phi_2= \phi\ ,
\ee
followed by the limit $N\to \infty$. Then, the new set of scalars obey
\be
\langle \phi_+(z) \phi_-(w)\rangle = -\langle \phi(z) \phi(w)\rangle =
\ln(z-w)\
\ee
and have zero correlators otherwise.
Following a procedure similar to section 3, we find that
\ba
&& \Psi=-{i\ov 2} (\phi^2+1)\del\phi_+ -\phi\del\phi + i \del\phi_-\ ,
\qq \Phi= -i\del\phi_+\ ,
\nonumber\\
&& P= {1\ov \sqrt{2}}(i\del\phi - \phi \del\phi_+)\ .
\label{JHWE}
\ea
Also, the stress-energy tensor of the theory becomes
\be
T(z)=-\ha (\del \phi)^2 + \del \phi_+\del\phi_- -{i\ov 2}\del^2 \phi_+\
\ee
and we thus complete the free field realization of the basic operators
in our logarithmic conformal field theory.

We conclude this section by presenting some ideas towards the construction
of more logarithmic partner fields in the model.
We may take advantage of the extended conformal symmetries of the
time-like free
boson model $U(1)_{-N}$ in order to define more fields of higher dimension,
which
become logarithmic partners in the large $N$ limit. For this purpose,
we use
the higher parafermion currents $\psi_{l}$ and
$\psi_l^{\dagger}$ to define the fields
\ba
\Phi_l(z) &=& \epsilon \left({\sqrt{N} \over 2} (\psi_l + \psi_l^{\dagger}) -
\sqrt{N} W_l \right) , ~~~~
\Psi_l(z) = {\sqrt{N} \over 2} (\psi_l + \psi_l^{\dagger}) +
\sqrt{N} W_l ~ , \nonumber\\
P_l(z) &=& \sqrt{\epsilon}{\sqrt{N} \over 2i}(\psi_l - \psi_l^{\dagger}) ~,
\ea
where $W_l$ are the chiral $W$-currents of the extra time-like boson.
They naturally generalize the fields that were defined earlier for $l=1$ to
operators of integer conformal dimension
$l \geq 2$ using the higher parafermions,  as one further takes the
limit $N \rightarrow \infty$.

Recall at this point that the conformal field theory
of a free boson admits a $W_{1 + \infty}$ symmetry\footnote{Actually,
there is a realization of the $W_{1+\infty}$
algebra in terms of a complex fermion,
which can be bosonized to yield a realization in terms of a real free scalar
field.}
with central charge $c=1$; it is
generated by the $U(1)$ current $J_0/\sqrt{N} = V_1$, the stress energy tensor
$T_{U(1)} = V_2$, and an infinite collection of higher spin fields
$V_l$ of integer
dimension $l=3, 4, 5, \cdots$. This algebra can be
twisted by the $U(1)$ current
of the model to yield a $W_{\infty}$ algebra generated by higher spin operators
$W_l$ for all $l\geq 2$. $W_2$ is the improved stress-energy tensor,
whereas the
remaining $W_l$ generators are also obtained by twisting the corresponding
$W_{1+\infty}$ generators $V_l$. The details of the algebra are not important for
our present purposes apart from the following points: first, the twisting allows
for the $U(1)$ current algebra to decouple from the remaining generators $W_l$;
it also amounts to changing the central charge of the Virasoro algebra from
$c=1$ to $c=-2$; finally, there is a quasi-primary basis for the $W_{\infty}$
generators in which the algebra linearizes and the 2-point function of all
generating fields becomes diagonal, i.e.,
\be
\langle W_l(z) W_{l^{\prime}}(w) \rangle = {c \over 4(z-w)^{l + l^{\prime}}}
\delta_{ll^{\prime}} ~; ~~~~~ {\rm with} ~ c=-2 ~,
\ee
where any spin dependent normalization factors have been absorbed into the
definition of the $W_l$ currents. Further
technical details on the subject of $W_{\infty}$
algebras can be found in \cite{baki,prs}, for instance,
and references therein.

It is then straightforward to verify, in analogy with the calculations that were
performed earlier, that the fields $\Phi_l$ and $\Psi_l$ have 2-point correlation
functions of the form
\ba
\langle\Psi_l(z) \Psi_{l^{\prime}}(w)\rangle & = &
l^2 {\ln(z-w) \over (z-w)^{l+l^{\prime}}}
\delta_{ll^{\prime}} ~, ~~~~
\langle\Phi_l (z) \Psi_{l^{\prime}}(w)\rangle =
{1 \over (z-w)^{l+l^{\prime}}}
\delta_{ll^{\prime}} ~, \nonumber\\
\langle \Phi_l(z) \Phi_{l^{\prime}}(w) \rangle & = & 0
\ea
as $N \rightarrow \infty$, for all values $l = 1, 2, 3, \cdots$.
Therefore, they
qualify to be logarithmic partners in the resulting LCFT.
However, it is fair to say that the fields $\Psi_l$ defined in this fashion
do not always have well-defined operator product expansion with the
ordinary stress-energy tensor of the theory, as central terms can appear
with a $\sqrt{N}$ dependence.


\section{Conclusions}
\setcounter{equation}{0}
Plane wave solutions have been studied extensively in the theory of
general relativity, but also in string theory, as they exhibit many
interesting physical properties. Their simplicity has also proven very useful
to address problems of gravitational physics within strings. In fact,
there have been some exactly solvable models based on non-semi-simple
groups that exhibit plane waves as supergravity backgrounds.

The interest in plane wave solutions arising in string and M-theory has been
recently revived following the realization that the original AdS/CFT
correspondence can be extended in a non-trivial way
to include the effect of highly massive string states
using such backgrounds \cite{BMN}.
The prototype example in these cases is the maximally supersymmetric
plane wave solution of type-IIB supergravity \cite{KoGli,Blau1}. This
solution, in turn, can be obtained as a Penrose limit of
the maximally supersymmetric
vacuum solution, $AdS_5 \times S^5$ \cite{Blau2}.
Note that, although the Penrose
limit was originally considered in general relativity,
it can be straightforwardly
extended to supergravities that correspond to low energy string theories
\cite{Gueven}.
Then, in accordance with experience and general expectations, the superalgebra
for the $AdS_5 \times S^5$ background contracts to the
corresponding plane wave
superalgebra, as it has been explicitly demonstrated in \cite{superal}.
Here, we have derived an unexpected result relating plane wave solutions
to an exact logarithmic conformal field theory.
Although there is no direct evidence that the logarithmic structure we
have encountered in the conformal field theory living on the world-sheet
has any relation to the logarithmic behaviour of correlators
in gauge theories (see, for instance, \cite{massimo}), we think that such
occurrence might not be accidental and worth exploring in the future.

The primary purpose of the present work was to understand the world-sheet
interpretation of the plane wave limit in supergravity, as we believe it
will sharpen our present understanding of all current issues that are
involved. We have
been able to report progress in this direction by considering a simple,
yet illustrative, example of an exact conformal field theory whose
Saletan limit (in the language of the world-sheet current algebra) gives
rise to plane waves in target space.
From the technical point of view,
we made extensive use of the parafermion
algebra, which is present in this representative model, to define operators
that are correlated to $U(1)$ currents and yield logarithmic correlation
functions in a suitable large $N$ limit.
Actually, our findings could have
been reported some years ago in the context of two-dimensional conformal
field theories, but there was no good reason to expect at
that time that there was a
correlated large $N$ limit, which made sense and could also
provide us with a totally different algebraic structure from
the ordinary situation. The emergence of
a logarithmic conformal field theory is quite interesting, as it can be
easily generalized to many more backgrounds having exact CFT description.
Moreover, our solution may be used to provide new examples of logarithmic
conformal field theories, which have been studied
so far only occasionally and in
a different context. Thus, it may put both subjects in a different
perspective and catalyze further the exchange of ideas among them.

There are certainly a number of questions that require further study.
The representation theory of the chiral algebras that arise in these
logarithmic conformal field theories is one of them and it has to be
developed in view of their applications in the gravitational plane wave
backgrounds. It might also be possible to obtain new results for the
generalized AdS/CFT correspondence, as well as explain several of the new
findings, by relying on the logarithmic structure that seems to exist
on the string world-sheet. Finally, more examples of logarithmic fields,
as well as a more systematic study of other coset models with plane wave limit
should be considered in detail. We hope to report on all these issues in
future publications.


\vspace{8 mm}

\centerline {\bf Acknowledgments}
\noindent
This work was supported in part by the European Research and Training Networks
``Superstring Theory" (HPRN-CT-2000-00122) and ``The Quantum Structure of
Space-time" (HPRN-CT-2000-00131). We also acknowledge support by the Greek State
Scholarships Foundation under the contract IKYDA-2001/22, as well as NATO support
by a Collaborative Linkage Grant under the contract PST.CLG.978785.



\newpage

\begin{thebibliography}{99}
\renewcommand{\baselinestretch}{1}
\normalsize


\bibitem{Witten}
E.~Witten,
Phys. Rev. {\bf D44} (1991) 314.

\bibitem{Horne}
J.H.~Horne and G.~T.~Horowitz,
Nucl. Phys. {\bf B368} (1992) 444,
{\tt hep-th/9108001}.

\bibitem{Bars}
I.~Bars and K.~Sfetsos,
Mod. Phys. Lett. {\bf A7} (1992) 1091
{\tt hep-th/9110054} and
Phys. Lett. {\bf B277} (1992) 269,
{\tt hep-th/9111040}.

\bibitem{Nappi}
C.R.~Nappi and E.~Witten,
Phys. Lett. {\bf B293} (1992) 309,
{\tt hep-th/9206078}.

\bibitem{Johnson}
C.V.~Johnson,
Phys. Rev. {\bf D50} (1994) 4032,
{\tt hep-th/9403192}.

\bibitem{nonco}
J.~Balog, L.O'Raifeartaigh, P.~Forgacs and A.~Wipf,
Nucl. Phys. {\bf B325} (1989) 225.

\bibitem{Petropoulos}
P.M.~Petropoulos,
Phys. Lett. {\bf B236} (1990) 151.

\bibitem{BarsNe}
I.~Bars and D.~Nemeschansky,
Nucl. Phys. {\bf B348} (1991) 89.

\bibitem{Gawedzki}
K.~Gawedzki,
{\it Noncompact WZW conformal field theories},
{\tt hep-th/9110076}.


\bibitem{NaWi}
C.R.~Nappi and E.~Witten,
Phys. Rev. Lett. {\bf 71} (1993) 3751,
{\tt hep-th/9310112}.


\bibitem{KiKouLu}
E.~Kiritsis and C.~Kounnas,
Phys. Lett. {\bf B320} (1994) 264
[Addendum-ibid. {\bf B325} (1994) 536],
{\tt hep-th/9310202};
E.~Kiritsis, C.~Kounnas and D.~Lust,
Phys. Lett. {\bf B331} (1994) 321,
{\tt hep-th/9404114}.

\bibitem{sfe1}
K.~Sfetsos,
Phys. Lett. {\bf B324} (1994) 335,
{\tt hep-th/9311010}.

\bibitem{Pen}
R. Penrose, {\it Any space-time has a wave as a limit}, Differential
Geometry and Relativity, Reidel, Dordrecht, 1976.

\bibitem{Gilmore}
R. Gilmore, {\it Lie Groups, Lie Algebras and Some of
Their Applications}, John Wiley.

\bibitem{Olive}
D.I.~Olive, E.~Rabinovici and A.~Schwimmer,
Phys. Lett. {\bf B321} (1994) 361,
{\tt hep-th/9311081}.

\bibitem{sfe2}
K.~Sfetsos,
Phys. Rev. {\bf D50} (1994) 2784,
{\tt hep-th/9402031}.

\bibitem{Mohammedi}
N.~Mohammedi,
Phys. Lett. {\bf B325} (1994) 371
{\tt hep-th/9312182}.

\bibitem{FF}
J.M.~Figueroa-O'Farrill and S.~Stanciu,
Phys. Lett. {\bf B327} (1994) 40,
{\tt hep-th/9402035}.

\bibitem{Forgacs}
P.~Forgacs, P.A.~Horvathy, Z.~Horvath and L.~Palla,
Heavy Ion Phys. {\bf 1} (1995) 65,
{\tt hep-th/9503222}.

\bibitem{Antoniadis}
I.~Antoniadis and N.~A.~Obers,
Nucl. Phys. {\bf B423} (1994) 639,
{\tt hep-th/9403191}.


\bibitem{sfetse}
K.~Sfetsos and A.A.~Tseytlin,
Nucl. Phys. {\bf B427} (1994) 245,
{\tt hep-th/9404063}.


\bibitem{saleur}
L.~Rozansky and H.~Saleur,
Nucl. Phys. {\bf B376} (1992) 461.


\bibitem{Gur}
V. Gurarie, Nucl. Phys. {\bf B410} (1993) 535,
{\tt hep-th/9303160}.



\bibitem{BK}
A. Bilal and I. Kogan, Nucl. Phys. {\bf B449} (1995) 569, {\tt hep-th/9503209}.

\bibitem{Cau}
J.-S. Caux, I. Kogan and A. Tsvelik, Nucl. Phys. {\bf B466}
(1996) 444, {\tt hep-th/9511134}.

\bibitem{KN}
I. Kogan and A. Nichols, JHEP {\bf 0201} (2002) 029,
{\tt hep-th/0112008}.

\bibitem{persis}
A.~M.~Ghezelbash, M.~Khorrami and A.~Aghamohammadi,
Int. J. Mod. Phys. {\bf A14} (1999) 2581,
{\tt hep-th/9807034}.

\bibitem{gardy}
J.L.~Cardy,
J. Phys. {\bf A25} (1992) L201, {\tt hep-th/9111026}.



\bibitem{saleur2}
H.~Saleur,
Nucl. Phys. {\bf B382} (1992) 486,
hep-th/9111007.
\bibitem{Flrev}
M. Flohr, {\it Bits and Pieces in Logarithmic Conformal Field
Theory}, {\tt hep-th/0111228}.

\bibitem{paraf}
A.B. Zamolodchikov and V.A. Fateev, Sov. Phys. JETP {\bf 62} (1985) 215.

\bibitem{claspara}
K.~Bardacki, M.J.~Crescimanno and E.~Rabinovici,
Nucl. Phys. {\bf B344} (1990) 344.

\bibitem{baki}
I. Bakas and E. Kiritsis, Nucl. Phys. {\bf B343} (1990) 185;
Prog. Theor. Phys. Suppl. {\bf 102} (1990) 15;
Int. J. Mod. Phys. {\bf A6} (1991) 2871;
Int. J. Mod. Phys. {\bf A7} (1992) 55.

\bibitem{prs}
C. Pope, L. Romans and X. Shen, Nucl. Phys. {\bf B339} (1990) 191;
Phys. Lett. {\bf B245} (1990) 72; E. Sezgin, {\it Aspects of $W_{\infty}$
Symmetry}, in the proceedings of the Tehran meeting on Mathematical
Physics (1990), {\tt hep-th/9112025}.


\bibitem{Amati}
D.~Amati and C.~Klimcik,
Phys. Lett. {\bf B219} (1989) 443.

\bibitem{Horowitz}
G.T.~Horowitz and A.R.~Steif,
Phys. Rev. Lett. {\bf 64} (1990) 260.


\bibitem{BMN}
D.~Berenstein, J.~Maldacena and H.~Nastase,
JHEP {\bf 0204} (2002) 013,
{\tt hep-th/0202021}.


\bibitem{KoGli}
J.~Kowalski-Glikman,
Phys. Lett. {\bf B134} (1984) 194.


\bibitem{Blau1}
M.~Blau, J.~Figueroa-O'Farrill, C.~Hull and G.~Papadopoulos,
{\it A new maximally supersymmetric background of IIB superstring theory},
JHEP {\bf 0201} (2002) 047,
{\tt hep-th/0110242}.

\bibitem{Blau2}
M.~Blau, J.~Figueroa-O'Farrill, C.~Hull and G.~Papadopoulos,
{\it Penrose limits and maximal supersymmetry},
{\tt hep-th/0201081} and
{\it Penrose limits, supergravity and brane dynamics},
{\tt hep-th/0202111}.


\bibitem{Gueven}
R.~Gueven,
Phys. Lett. {\bf B482} (2000) 255,
{hep-th/0005061}.


\bibitem{superal}
M.~Hatsuda, K.~Kamimura and M.~Sakaguchi,
{\it From super-$AdS(5) \times S^5$ algebra to super-pp-wave algebra},
{\tt hep-th/0202190}.

\bibitem{massimo}
M.~Bianchi, S.~Kovacs, G.~Rossi and Y.S.~Stanev,
JHEP {\bf 9908} (1999) 020,
{\tt hep-th/9906188}.


\end{thebibliography}
\end{document}